\documentclass[conference]{IEEEtran}
\IEEEoverridecommandlockouts
\usepackage{cite}
\usepackage{amsmath,amssymb,amsfonts}
\usepackage{algorithmic}
\usepackage{graphicx}
\usepackage{textcomp}
\usepackage{multirow}
\usepackage{xcolor}
\usepackage{array}
\usepackage{bm}
\usepackage{amsmath}
\usepackage{booktabs} 
\def\BibTeX{{\rm B\kern-.05em{\sc i\kern-.025em b}\kern-.08em
    T\kern-.1667em\lower.7ex\hbox{E}\kern-.125emX}}
\begin{document}
\setlength{\jot}{1pt}
\title{
Two-Stage Adaptive Robust Optimization Model for \\Joint Unit Maintenance and Unit Commitment \\Considering Source-Load Uncertainty
\\
\thanks{This work was supported by the National Natural Science Foundation of China (52307137, 52222705). Corresponding author: Yuxiong Huang.}
}


\author{Hongrui Lu, Yuxiong Huang, Tong He, Gengfeng Li\\
\textit{School of Electrical Engineering, Xi’an Jiaotong University}, Xi'an, 710049, China  \\
luhongrui@stu.xjtu.edu.cn,  yuxionghuang@xjtu.edu.cn, t-he23@stu.xjtu.edu.cn, gengfengli@xjtu.edu.cn
}

\maketitle

\begin{abstract}
Unit maintenance and unit commitment are two critical and interrelated aspects of electric power system operation, both of which face the challenge of coordinating efforts to enhance reliability and economic performance. This challenge becomes increasingly pronounced in the context of increased integration of renewable energy and flexible loads, such as wind power and electric vehicles, into the power system, where high uncertainty is prevalent. To tackle this issue, this paper develops a two-stage adaptive robust optimization model for the joint unit maintenance and unit commitment strategy. The first stage focuses on making joint decisions regarding unit maintenance and unit commitment, while the second stage addresses economic dispatch under the worst-case scenarios of wind power and load demand. Then a practical solution methodology is proposed to solve this model efficiently, which combines the inexact column-and-constraint generation algorithm with an outer approximation method. Finally, the economic viability and adaptability of the proposed method is demonstrated based on the RTS-79 test system.


\end{abstract}


\section{Introduction}
The increasing uncertainties of the renewable power system create greater demands for the coordination of unit maintenance (UM) and unit commitment (UC), to improve power supply reliability. Traditionally, UM strategies aim to minimize the risk of power loss or economic costs to adjust the reporting schedule \cite{lu1, he1}, while UC strategies are arranged regarding UM as the boundary conditions \cite{lu3} . However, decoupling UM and UC may lead to an increasing risk of power loss and a higher operation cost. Therefore, some scholars have conducted research on the joint UM and UC (J-UMUC). Ref. \cite{lu2} develops an integrated optimization model that coordinately optimizes UM and UC in the context of asset loading. Ref. \cite{lu4} formulates a computationally efficient method for solving large instances of the J-UMUC problem. Ref. \cite{lu9} coordinates short-term with medium-term UM, considering short-term security-constrained UC. However, the above studies ignore the uncertainty of the source and load, which is not suitable for the renewable power system. Thus, it is necessary to conduct the J-UMUC research considering the uncertainty of source and load.

The main current methods for solving uncertainty optimization problems are stochastic optimization (SO) and robust optimization (RO). Unlike SO requires the probability distributions of uncertain scenarios, which is difficult to obtain, RO is committed to finding the worst-case scenario and provides a risk-averse solution, which has been widely applied in the power system \cite{robust}. As effective solution methods for RO, Benders decomposition generates a large number of cutting planes at each iteration, but may lead to high computational complexity, and column-and-constraint generation ($\text{C\&CG}$) dynamically generates variables and constraints, but the computational speed decreases with the iterative process. Therefore, Ref. \cite{lu15} proposed an inexact column-and-constraint generation ($\text{i-C\&CG}$) algorithm and demonstrates its computational advantages through numerical
experiments.

In this paper, a two-stage adaptive robust optimization model is proposed for the J-UMUC strategy considering source and load uncertainty, where the first-stage J-UMUC decisions and the second-stage dispatch decisions are robust under source and load uncertainty. Since the model has a “min-max-min” structure, which cannot be solved directly, we decouple the original problem into the master problem and the sub-problem, then transform the bi-lever sub-problem into a single-lever problem by strong duality, which we solve using the $\text{i-C\&CG}$ algorithm. Finally we verify the feasibility and effectiveness of the proposed model as well as the efficiency advantage of the algorithm. The main contributions in this paper are as follows: 

1) A two-stage adaptive robust optimization model is proposed for the J-UMUC strategy considering source and load uncertainty, which can satisfy the needs of both UM and UC with economic efficiency and adaptability. 

2) A practical solution methodology to solve the adaptive robust model is proposed, where an $\text{i-C\&CG}$ algorithm is employed to improve the computational efficiency and the inner sub-problems with the bi-linear terms are solved by an outer approximation (OA) approach \cite{lu17}. 

3) The feasibility and effectiveness of the proposed model is verified and the impact of different algorithm parameters on the solution speed has been analyzed.
 
\vspace{-20pt}
\section{Model Formulation}

\subsection{Uncertainty Set}
Since that the Box Uncertainty Set is limited by essential information about the variable distribution, resulting in relatively high conservatism \cite{robust}, we construct an uncertainty set $\boldsymbol{\mathcal{V}}$ of the following form:
\begin{align}
\boldsymbol{\mathcal{V}} = \{\bm{\tilde{D}}\in  \mathcal{R}^{N_{\mathrm{D}} \times T},\bm{\tilde{P}^{\mathrm{W}}}\in  \mathcal{R}^{N_{\mathrm{W}} \times T} : \notag\\
\tilde{D}_{i, t} \in [\overline{D}_{i, t} - \hat{D}_{i, t}  , \overline{D}_{i, t} + \hat{D}_{i, t}] ,
\sum_{i=1}^{N_{\mathrm{D}}} \left| \frac{\tilde{D}_{i, t} - \overline{D}_{i, t}} {\hat{D}_{i, t}} \right| \leq \varGamma_{\mathrm{D}} \notag\\
  \tilde{p}_{w,t}^\mathrm{W} \in [\overline{p}_{w,t}^\mathrm{W} - \hat{p}_{w,t}^\mathrm{W} , \overline{p}_{w,t}^\mathrm{W} + \hat{p}_{w,t}^\mathrm{W}],
\sum_{i=1}^{N_{\mathrm{W}}} \left| \frac{\tilde{p}_{w,t}^\mathrm{W} - \overline{p}_{w,t}^\mathrm{W}} {\hat{p}_{w,t}^\mathrm{W}} \right| \leq \varGamma_{\mathrm{W}}\notag\\
\forall t=1,2,...,T,
\forall i=1,2,...,N_{\mathrm{D}},
\forall w=1,2,...,N_{\mathrm{W}}\}
\label{1}
\end{align}
where $N_\mathrm{D}$ and $N_{\mathrm{W}}$ represent the total number of loads and wind turbines, respectively; $T$ denotes the number of time periods; $\overline{D}_{i, t}$ and $\overline{p}_{w, t}^\mathrm{W}$ denote the forecast value of the $i$-th load and the $w$-th wind power output at time $t$; $\hat{D}_{i, t}^{+}$ and $\hat{D}_{i, t}^{-}$ denote the upper and lower bounds of the forecast error of the $i$-th load at time $t$; $\hat{p}_{w, t}^{\mathrm{W}+}$ and $\hat{p}_{w, t}^{\mathrm{W}-}$ denote the upper and lower bounds of the forecast error of the $w$-th wind power output at time $t$; $\varGamma_{\mathrm{D}}$ and $\varGamma_{\mathrm{W}}$ are the uncertainty budget parameters of the load and wind power output, determining the size of the uncertainty set. As $\varGamma_{\mathrm{D}}$ or $\varGamma_{\mathrm{W}}$ increases, the size of uncertainty set also expands. Specifically, if  $\varGamma_{\mathrm{D}}=\varGamma_{\mathrm{W}}=0$, the uncertainty set reduces to a singleton.

\subsection{Two-Stage Robust Optimization Model}
The proposed robust optimization method is divided into two stages. In the first stage, the J-UMUC strategy is formulated, while in the second stage, economic dispatch is conducted under the worst-case scenario from the uncertainty set, based on the decisions made in the first stage. The objective function of the two-stage robust optimization model is as follows:
\begin{equation}
\left \{
\begin{array}{ll}
\mathop {\min }\limits_{\textbf{q},\textbf{u}} F_{1} + \mathop {\max }\limits_{\bm{v} \in \boldsymbol{\mathcal{V}}} \mathop {\min }\limits_{\left\{ {\bm{\Delta p^{\text{W}}}, \textbf{d}, \textbf{f}, \textbf{p}, \bm{\delta}} \right\}} F_2\\
F_{1} = \sum\limits_{n = 1}^{N_\mathrm{M}}  \sum\limits_{t = 1}^{T} C_{n,t}^\mathrm{M} q_{n,t} + \sum\limits_{g = 1}^{N_\mathrm{G}}  \sum\limits_{t = 1}^{T} ( C_{g,t}^\mathrm{S} + C_{g}^\mathrm{NL} u_{g,t}) \\
F_{2} = \sum\limits_{t = 1}^{T}(\sum\limits_{i = 1}^{N_\mathrm{D}}   C^\mathrm{LS} d_{i,t} + \sum\limits_{g = 1}^{N_\mathrm{G}}   C_{g}^\mathrm{G}u_{g,t}(p_{g,t}-p_{g,\mathop {\min }})&\\ \ \ \ \ \ \ \ \ \ \ \ +\sum\limits_{w = 1}^{N_\mathrm{W}}   C^\mathrm{WS} \Delta p_{w,t}^\mathrm{W})&\\
\end{array}
\right.
\label{2}
\end{equation}
where $F_1$ represents the cost of the J-UMUC strategy; $F_2$ represents the costs of economic dispatch and penalties for load shedding and wind power curtailment; $N_{\mathrm{M}}$ is the number of units awaiting maintenance; $C_{n,t}^{\mathrm{M}}$ is the maintenance cost of unit $n$ in period $t$; $q_{n,t}$ is a binary variable indicating whether unit $n$ begins maintenance at time $t$ (1 if begin, 0 otherwise); $u_{g,t}$ is a binary variable indicating the commitment status of the $g$-th unit at time $t$ (1 if on-line, 0 otherwise); $N_{\mathrm{G}}$ is the number of conventional units; $C_{g,t}^{\mathrm{S}}$ is the start-up cost of unit $g$ at time $t$; $C_{g}^{\mathrm{NL}}$ is the cost generated by unit $g$ at its minimum technical output; $C^{\mathrm{LS}}$ and $C^{\mathrm{WS}}$ are the penalty costs for load shedding and wind power curtailment; $d_{i,t}$ is the load shedding of the $i$-th load  at time $t$; $C_{g}^{\mathrm{G}}$ is the cost per unit for the output of unit $g$ exceeding its minimum technical output; $p_{g,\min}$ is the minimum technical output of unit $g$.

The constraints are as follows:
\subsubsection{Unit maintenance constraints}
\begin{align}
   q_{n,t} \geq (m_{n,t} -m_{n,t-1}),q_{n,1} \geq m_{n,1},
   \forall t=2,...,T,\forall n 
\label{3}
\end{align}
\vspace{-20pt}
\begin{align}
m_{n,t} -m_{n,t-1} \leq m_{n,t-1 +S_{n}} ,m_{n,1}\leq m_{n,S_{n}} \notag\\
\forall t=2,...,T+1-S_{n},\forall n 
\label{4}
\end{align}
\vspace{-15pt}
\begin{equation}
\sum\limits_{t=1}^{T} m_{n,t} = S_{n},\sum\limits_{t=1}^{T} q_{n,t} = 1 ,\forall n
\label{5}
\end{equation}
\vspace{-5pt}
\begin{equation}
\sum\limits_{n=1}^{N_\mathrm{M}} m_{n,t} \leq R_{t} ,\forall t
\label{6}
\end{equation}
where $m_{n,t}$ is a binary variable indicating the maintenance status of unit $n$ at time $t$ (1 if under maintenance, 0 otherwise); $S_{n}$ is the maintenance time of unit $n$; $R_{t}$ is the maintenance resource budget at time $t$. Constraint (\ref{3}) link the maintenance start indicators with the maintenance status, (\ref{4}) ensure maintenance continuity, (\ref{5}) restricts the limit of maintenance duration and number, and (\ref{6}) represents the maintenance resource budget.

\subsubsection{Unit commitment constraints}
\begin{align}
       \sum\limits_{k=t}^{t+UT_{g}-1}{\left( 1-{u}_{g,k} \right)}\ge UT_{g}\left( {{u}_{g,t-1}}-{{u}_{g,t}} \right),\notag\\ \forall t=2,...,(T-(UT_{g}-1)),\forall g  
\label{7}
\end{align}
\vspace{-22pt}
\begin{align}
    \sum\limits_{k=t}^{t+DT_{g}-1} {u}_{g,k} \ge DT_{g}\left( {{u}_{g,t}}-{{u}_{g,t-1}} \right),\notag\\ \forall t=2,...,(T-(DT_{g}-1)),\forall g 
    \label{8}
\end{align}
\vspace{-15pt}
\begin{equation}
\sum\limits_{g=1}^{N_\mathrm{G}} u_{g,t}p_{g,\mathop{\max}} \geq  \rho (\sum\limits_{i=1}^{N_\mathrm{B}}\overline{D}_{i, t}-\sum\limits_{w=1}^{N_\mathrm{W}}\overline{p}_{w, t}^\mathrm{W}) ,\forall t
\label{9}
\end{equation}
\vspace{-10pt}
\begin{equation}
 C_{g,t}^\mathrm{S} = \tau_{g,t} C_{g}^\mathrm{S},\tau_{g,t} \geq u_{g,t}-u_{g,t-1},\tau_{g,1} \geq u_{g,1},\forall t,g
 \label{10}
\end{equation}
\vspace{-10pt}
\begin{equation}
-M(1-m_{n,t}) \leq u_{n,t} \leq M(1-m_{n,t}) ,\forall t, n
\label{11}
\end{equation}
\vspace{-10pt}
\begin{equation}
    u_{g,t} p_{g,\mathop{\min}} \leq p_{g,t} \leq u_{g,t} p_{g,\mathop{\max}} ,\forall t, g
    \label{12}
\end{equation}
\vspace{-10pt}
\begin{equation}
 \left \{
\begin{array}{ll}    
p_{g,t} - p_{g,t-1} \leq u_{g,t-1} RU_{g} + (1-u_{g,t-1}) RSU_{g} ,\\
p_{g,t-1} - p_{g,t} \leq u_{g,t}RD_{g} +(1-u_{g,t})RSD_{g},\\
p_{g,1} \leq RSU_{g},\forall t=2,...,T,\forall g 
\end{array}
\right.
\label{13}
\end{equation}
where $UT_{g}$ and $DT_{g}$ are the minimum on-time and off-time of unit $g$; $\rho$ is the system hot standby rate;  $\tau_{g,t}$ is a binary variable indicating the start-up status of unit $g$ at time $t$ (1 if started-up, 0 otherwise); $p_{g,\mathop{\max}}$ is the maximum output of unit $g$; $RU_{g}$ and $RD_{g}$ are the ramp-up and ramp-down rate of unit $g$; $RSU_{g}$ and $RSD_{g}$ are the maximum start-up and shut-down rate of unit $g$. Constraints (\ref{7}) and (\ref{8}) are the start-up and shut-down time constraints, (\ref{9}) describes reserve requirements, and (\ref{10}) represents the start-up cost of unit $g$. In (\ref{11}), $M$ is a sufficiently positive number, it indicates that the units under maintenance cannot be on-line. (\ref{12}) enforces the upper and lower limits of the output, while (\ref{13}) establishes the ramp rate limits.

\subsubsection{Economic dispatch constraints}
\begin{equation}
-F_{l} \le {f_{l,t}} \le F_{l},\forall t,l
\label{14}
\end{equation}
\vspace{-15pt}
\begin{equation}
-\overline{\delta}  \le {\delta _{i,t}} \le \overline{\delta} ,\forall t,i
\label{15}
\end{equation}
\vspace{-15pt}
\begin{equation}
0 \le {d_{i,t}} \le \tilde{D}_{i, t},\forall t,i
\label{16}
\end{equation}
\vspace{-15pt}
\begin{equation}
0 \leq \Delta p_{w,t}^\mathrm{W}\leq \tilde{p}_{w,t}^\mathrm{W},\forall t, w
\label{17}
\end{equation}
\vspace{-15pt}
\begin{equation}
f_{l,t}x_l = {\delta _{o(l),t}} - {\delta _{d(l),t}},\forall t,l
\label{18}
\end{equation}
\vspace{-15pt}
\begin{equation}
    \sum\limits_{w\in {\boldsymbol{\mathcal{W}_i}}}(\tilde{p}_{w,t}^\mathrm{W}- \Delta {p}_{w,t}^\mathrm{W}) + \sum\limits_{g \in {\boldsymbol{\mathcal{G}_i}}} p_{g,t}  - \sum\limits_{l\left| {o(l) = i} \right.} {f_{l,t}} \notag
\end{equation}
\vspace{-10pt}
\begin{equation}
    + \sum\limits_{l\left| {d(l) = i} \right.} {f_{l,t}}  + d_{i,t} = \tilde{D}_{i, t},\forall t,i
    \label{19}
\end{equation}
where $f_{l,t}$ is active power flow on line $l$ at time $t$; $\boldsymbol{\mathcal{W}_i}$ and $\boldsymbol{\mathcal{G}_i}$ represent the sets of wind turbines and conventional units connected to node $i$, respectively; $o(l)$ and $d(l)$ are the parent and son node of line $l$; $x_l$ is the reactance of line $l$; $\delta_{i,t}$ is the voltage angle of bus $i$ at time $t$; $F_{l}$ is the limit of active power flow of line $l$; $\overline{\delta}$ is the limit of the voltage angle of bus $i$. Constraints (\ref{14}) and (\ref{15}) ensure that the power flow and voltage angle are within limits, (\ref{16}) and (\ref{17}) enforce the limits of load shedding and wind power curtailment, respectively, (\ref{18}) and (\ref{19}) ensure power balance at each bus using the DC power flow model.

\subsection{Model Reformulation}
Denote the scheduling variables in the first-stage as $\bm{x}$ and the dispatch variables in the second-stage as $\bm{y}$, we represent the proposed two-stage robust optimization model in the following matrix form:
\begin{align}
    \mathop {\min } \limits_{\bm{x}}\ & \bm{c}^{T}\bm{x} + \mathop {\max }\limits_{\bm{v} \in \boldsymbol{\mathcal{V}}} \mathop {\min }\limits_{\bm{y}}\ (\bm{b}^{T}\bm{y}+\bm{x}^{T}\bm{L}\bm{y} +\bm{k}^{T}\bm{x}) \notag\\
    \text{s.t.}\  &\bm{A}\bm{x} \geq \bm{e},\label{20} \\&
    \bm{E}\bm{x}+\bm{H}\bm{y} +\bm{G}\bm{v} \geq \bm{g},\label{21}\\&  \bm{F}\bm{y}+\bm{M}\bm{v}=\bm{f},\label{22}\\&
    \bm{x}\  \text{binary},\ \bm{y}\  \text{free}
\end{align}

Constraint (\ref{20}) corresponds to (\ref{3})-(\ref{11}), (\ref{21}) corresponds to (\ref{1}), (\ref{12})-(\ref{17}), and (\ref{22}) corresponds to (\ref{18})-(\ref{19}). \(\bm{A}\), \(\bm{E}\), \(\bm{H}\), \(\bm{G}\), \(\bm{F}\), \(\bm{M}\), \(\bm{L}\) are the corresponding coefficient matrices, and \(\bm{e}\), \(\bm{g}\), \(\bm{f}\), \(\bm{c}\), \(\bm{b}\), \(\bm{k}\) are constant column vectors.

It is evident that this two-stage robust optimization model has a “min-max-min” tri-level structure, making it unsolvable directly. In the second stage, $\bm{x}$ and $\bm{v}$ are given variables, so the third level is actually a linear problem. By applying strong duality, we can write the dual of the third level problem, transforming the middle and inner levels into a single maximization problem. Let $S(\bm{x},\bm{v})$ denote the objective function:
\begin{align}
   S(\bm{x},\bm{v})= \mathop {\min }\bm{c}^{T}&\bm{x}+\mathop {\max }\limits_{\bm{\lambda} ,\bm{\mu}} (\bm{\lambda}^{T} (\bm{g} - \bm{E}\bm{x}- \bm{G}\bm{v})\notag\\
    &\ \ \ \ +\bm{\mu}^{T} (\bm{f}-\bm{M}\bm{v} )+\bm{k}^{T}\bm{x})\label{24}\\ 
    \text{s.t.}\ & \bm{A}\bm{x} \geq \bm{e},\label{25}\\&   \bm{\lambda}^{T}\bm{H}+\bm{\mu}^{T}\bm{F}=\bm{b}^{T}+\bm{x}^{T}\bm{L},\label{26}\\&
    \bm{x}\  \text{binary},\bm{v} \in \mathcal{V},\bm{\lambda} \geq0,\bm{\mu}\ \text{free}
    \label{27}
\end{align}
where $\bm{\lambda}$,$\bm{\mu}$ are the dual variables for (\ref{21}) and (\ref{22}).

In this way, the original tri-level model is reformulated as an equivalent bi-level model.

\section{Solution Methodology}
In this section, we decouple the bi-level model into a master problem and a sub-problem, which we solve using the $\text{i-C\&CG}$ algorithm. The $\text{i-C\&CG}$ algorithm allows for an inexact solution to the master problem. By introducing an inexact relative gap as a buffer layer, the algorithm progresses through various steps until it converges. The detailed master problem, sub-problem, and overall algorithm procedure are presented below.

\subsection{The Master Problem}
Given a subset of temporal wind power and load values $\hat{\bm{V}}=\{\hat{\bm{v}}^{1},...,\hat{\bm{v}}^{k}\} \subseteq \boldsymbol{\mathcal{V}} $, we solve the master problem in the first stage to obtain the optimal J-UMUC strategy. For a specific uncertain scenario $\{\hat{\bm{v}}^{s}, s \leq k\}$, we define a set of second-stage variables $\bm{y}^{s}$. The master problem is then constructed as follows:
\begin{align}
   M(\bm{x})=& \mathop {\min } \limits_{\bm{x}}\  \bm{c}^{T}\bm{x} + \alpha \notag\\
    \text{s.t.} \ &\bm{A}\bm{x} \geq \bm{e}\\
    \label{29} &\alpha \geq  \bm{b}^{T}\bm{y}^{s}+\bm{x}^{T}\bm{L}\bm{y}^{s} +\bm{k}^{T}\bm{x},\forall s \leq k \\
    &\bm{E}\bm{x}+\bm{H}\bm{y}^{s} \geq \bm{g}-\bm{G}\hat{\bm{v}}^{s} ,\forall s \leq k\\
    &\bm{F}\bm{y}^{s}=\bm{f}-\bm{M}\hat{\bm{v}}^{s},\forall s \leq k\\
    &\bm{c}^{T}\bm{x} + \alpha \geq \overline{L}\\
    &\bm{x}\  \text{binary},\ \bm{y}^{s} \ \text{free},\ \forall s \leq k
\end{align}
where $\overline{L}$ is the lower bound of the master problem that updates with each iteration, as introduced by the $\text{i-C\&CG}$ algorithm. Note that (\ref{29}) is a nonlinear but can be linearized using the big-M method. For example, the bi-linear terms $u_{g,t} p^{s}_{g,t}$ in (\ref{29}) are replaced by $\eta^{s}_{g,t}$, with the additional constraints introduced as follows:
\begin{align}
-M(1-u_{g,t}) \leq \eta^{s}_{g,t}&-p^{s}_{g,t} \leq M(1-u_{g,t})\notag\\
-M u_{g,t} \leq &\eta^{s}_{g,t}\leq M u_{g,t}
\end{align}

Since the master problem is a single minimization problem and only considers a subset of the stochastic scenarios, which corresponds to relaxing the constraints, it provides a lower bound on the global optimal solution. By solving the master problem with a relative gap $\epsilon_{MP}$, we obtain the best solution $\hat{\bm{x}}$ and then record an upper bound $U_{in}=M(\hat{\bm{x}})$ and a lower bound $L_{in}\geq \overline{L}$.

\subsection{The Sub-Problem}
 The sub-problem aims to identify the worst-case uncertainty scenarios, which is equivalent to solving (\ref{24})-(\ref{27}) with ${\bm{x}}$ fixed at $\hat{\bm{x}}$. To address the bilinear terms $\bm{\lambda}^{T}\bm{G}\bm{v}$ and $\bm{\mu}^{T}\bm{M}\bm{v}$, we employ an outer approximation (OA) algorithm. The detailed steps of the OA algorithm are presented below:

\begin{table}[htbp]
\centering
\label{tab:example}
\begin{tabular*}{\linewidth}{@{\extracolsep{\fill}} l}
\toprule
\textbf{Algorithm 1} Outer Algorithm Solution Method \\
\midrule
\textbf{Initialization}: Set an initial uncertain scenario ${\bm{v}^{1}} \in \boldsymbol{\mathcal{V}}$, the OA lower bound \\$L_{OA}\leftarrow 0$, the OA upper bound $U_{OA}\leftarrow\infty$, iteration number  $j\leftarrow1$ and \\ the tolerance gap $\delta'$.\\
\textbf{while} $U_{OA}-L_{OA} \ge \delta'$ \textbf{do} \\
1. Solve OA sub-problem $S(\hat{\bm{x}},\bm{v}^{j})$, defined by (\ref{24})-(\ref{27}). Obtain $OASPobj$ \\and   the optimal 
solution $(\bm{\lambda}^{j},\bm{\mu}^{j})$, update $L_{OA} \leftarrow OASPobj$. Denote the \\linearization 
of bilinear terms $(\bm{\lambda}^{T}\bm{G}+\bm{\mu}^{T}\bm{M})\bm{v}$ at $(\bm{v}_j,\bm{\lambda}_j,\bm{\mu}_j)$ 
as:\\
\ \ \ \ \ \ \  $ L_{j}(\bm{v},\bm{\lambda},\bm{\mu})=(\bm{\lambda}_{j}^{T}\bm{G}+\bm{\mu}_{j}^{T}\bm{M})^{T}\bm{v}_{j}+(\bm{v}-\bm{v}_{j})^{T}(\bm{\lambda}_{j}^{T}\bm{G}+\bm{\mu}_{j}^{T}\bm{M})$\\
\ \ \ \ \ \ \ \ \ \ \ \ \ \ \ \  $+(\bm{\lambda}^{T}\bm{G}+\bm{\mu}^{T}\bm{M}-(\bm{\lambda}_{j}^{T}\bm{G}+\bm{\mu}_{j}^{T}\bm{M}))^{T}\bm{v}_{j}$\\
2. Solve OA master problem: \\
\ \ \ \ \ \ \ \ \ \ \ \ \ \ \ \ $\bm{c}^{T}\hat{\bm{x}}+\mathop {\max }\limits_{\bm{\lambda} ,\bm{\mu}} (\bm{\lambda}^{T} (\bm{g} - \bm{E}\hat{\bm{x}})
    +\bm{\mu}^{T} \bm{f}+\bm{k}^{T}\hat{\bm{x}})+\beta$\\
\ \ \ \ \ \ \ \ \ \ \ \ \ \ \ \ \ \ \ \ \ \ \ \ $
    \text{s.t.}\ \ \  \beta \leq L_{i}(\bm{v},\bm{\lambda},\bm{\mu})\  \forall i=1,...,j $\\
\ \ \ \ \ \ \ \ \ \ \ \ \ \ \ \ \ \ \ \ \ \ \ \ \ \ \ \ \ \ $ \bm{\lambda}^{T}\bm{H}+\bm{\mu}^{T}\bm{F}=\bm{b}^{T}+\hat{\bm{x}}^{T}\bm{L},$\\
\ \ \ \ \ \ \ \ \ \ \ \ \ \ \ \ \ \ \ \ \ \ \ \ \ \ \ \ \ \ $\bm{v} \in \mathcal{V},\ \bm{\lambda} \geq0,\ \bm{\mu}\ \text{free}$\\
Obtain $OAMPobj$ and the best solution $\bm{v}$, update $U_{OA} \leftarrow OAMPobj$ ,\\
update $U_{OA} -L_{OA}$ \\
3. $j\leftarrow j+1$, $\bm{v}^{j}\leftarrow\bm{v}$\\
\textbf{end while} \\
\textbf{return} $\overline{U} \leftarrow U_{OA}, \hat{\bm{v}}\leftarrow \bm{v}^{j}$\\
\bottomrule
\end{tabular*}
\end{table}

Considering that the sub-problem is equivalent to splitting the bi-level problem into two separate single-level optimization problems and summing the objective functions, it does not necessarily yield an optimal solution. Thus the sub-problem provides an upper bound $\overline{U}$.
\vspace{-10pt}
\subsection{Solution Algorithm}
\begin{figure}[htbp]
\centerline{\includegraphics[width=3.6in]{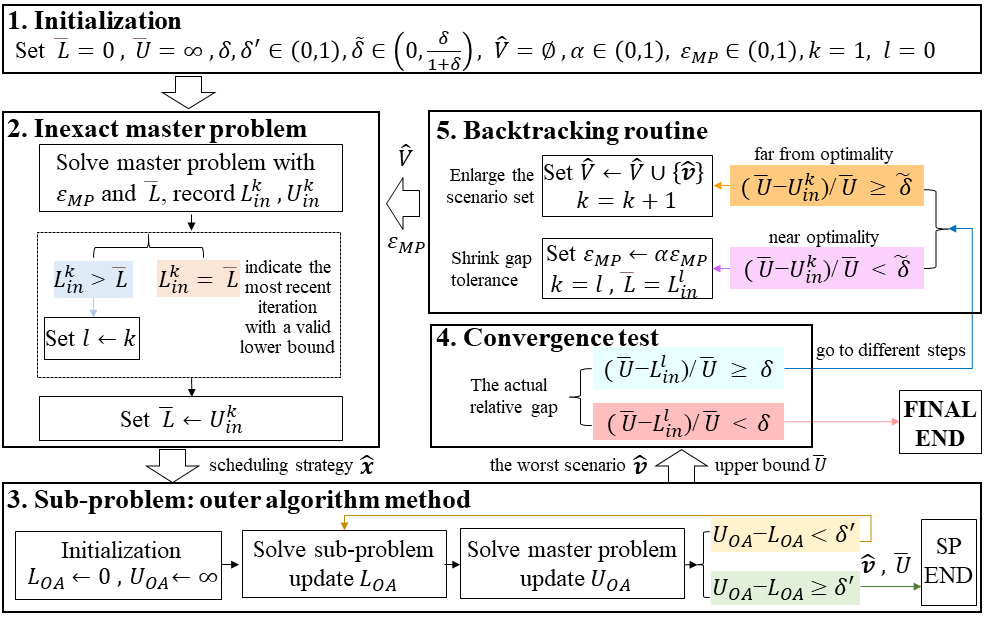}}
\vspace{-5pt}
\caption{ The illustration of the overall solution algorithm.}
\label{fig1}
\end{figure}

The flow chart of the overall solution algorithm is shown in Fig. \ref{fig1}, where $\delta$ is the relative optimization gap tolerance; $\tilde{\delta}$ is the inexact relative gap; $\alpha$ is the factor controlling the rate of descent of $\epsilon_{MP}$. Unlike the traditional $\text{C\&CG}$ method, the initial $\epsilon_{MP}$ is relatively large, leading to an inexact but fast solution to the master problem. In fact, the subsequent backtracking routine balances the imprecision. If the actual relative gap $(\overline{U}-L_{in})/\overline{U}$ does not satisfy convergence, the algorithm goes to a different step based on the inexact relative gap $(\overline{U}-U_{in})/\overline{U}$. If the inexact relative gap is less then $\tilde{\delta}$, that means the algorithm is nearing optimality, prompting a reduction in $\epsilon_{MP}$ to improve the accuracy of the master problem. Otherwise, we need to enlarge the scenario set with $\bm{\hat{v}}$ from sub-problems, as the current solution remains distant from the optimal solution.

It is worth emphasizing that all master problems and sub-problems are mixed-integer linear programs after linearization, allowing for efficient resolution using specialized mixed-integer programming solvers.

\vspace{-10pt}

\vspace{-20pt}
\section{Case Study}
The IEEE Reliability Test System (RTS-79) \cite{lu16} is used to verify the validity and feasibility of the proposed method. This system comprises 24 buses, 38 lines, 32 units, and 17 loads. Weekly load and unit data are sourced from \cite{lu16}, with a maximum load of 2,800 MW and unit ramp-up/ramp-down rates set at 40\% of their capacity. The system includes three wind turbines with capacities of 100, 200, and 200 MW, connected to buses 1, 3, and 9. The forecast error for wind power and load is set at 10\%, and the penalty costs for load shedding and wind power curtailment are \$3,000 and \$300 per kWh. The parameters for the units to be maintained are detailed in Table \ref{tab1}. The algorithm is implemented in MATLAB with Gurobi 10.0.0, running on a computer with an Intel i7-13620H CPU @ 2.40 GHz and 16 GB of RAM. The relative gap tolerance $\delta$ is set at 0.2\%, the OA gap tolerance $\delta'$ is set at 0.1\%, the inexact relative gap $\tilde{\delta}$ is set at 0.15\%, and the initial master relative gap $\epsilon_{MP}$ is set at 0.1\%.

\begin{table}[b]
\vspace{-20pt}
\caption{The parameters of the units to be maintained}
\vspace{-20pt}
\begin{center}
\begin{tabular}{|c|c|c|c|c|c|}
\hline
\multirow{2}*{\textbf{Unit}}& \textbf{Capacity}&\textbf{Start time} & \textbf{Initial} & \textbf{Penalty cost}& \textbf{ Duration}\\
~& \textbf{(MW)}&\textbf{(h)} & \textbf{cost (\$)} &\textbf{(\$/h)} & \textbf{(h)}\\
\hline
\textbf{G8} & 76 & 5 & 1,500 & 30 & 12\\
\hline
\textbf{G10} & 100 &  50 & 1,500 & 30 & 12\\
\hline
\textbf{G24} & 400 &  130 & 3,000 & 60 & 24\\
\hline
\end{tabular}
\label{tab1}
\end{center}
\vspace{-10pt}
\end{table}

\vspace{-5pt}
\subsection{Cost Efficiency and Adaptability}
To illustrate the effectiveness of the J-UMUC strategy under source-load uncertainty, we establish the following four cases for analysis:
\begin{itemize}
    \item Case 1: Not uncertainty is considered, and UM and UC are decoupled. The UM schedule is fixed at the reported time, followed by the UC schedule.    
    \item Case 2: Not uncertainty is considered, but UM and UC are coordinated. The J-UMUC strategy is based on the forecast values of load and wind power output.    
    \item Case 3: Uncertainty is considered, but UM and UC are decoupled. The uncertainty budgets are set at $\varGamma_{\mathrm{W}}=0.2 N_{\mathrm{W}},\varGamma_{\mathrm{D}}=0.2 N_{\mathrm{D}}$.
    \item Case 4: Uncertainty is considered, and UM and UC are coordinated, representing our proposed method. The uncertainty budgets are the same as in Case 3.
\end{itemize}

\begin{table}[b]
\vspace{-10pt}
\caption{Comparison of results under 4 cases}
\vspace{-5pt}
\centering
\begin{tabular}{|c|c|c|c|c|c|}
\hline
\multicolumn{2}{|c|}{\textbf{Case}}&\textbf{1}&\textbf{2}&\textbf{3}&\textbf{4}\\
\hline
\multicolumn{2}{|c|}{\textbf{Maintenance Cost (k\$)}} & 6 & 6.9 & 6 & 6.39\\
\hline
\multicolumn{2}{|c|}{\textbf{Commitment Cost (k\$)}} &  738.62 & 738.50& 779.08 &775.16\\
\hline
\multicolumn{2}{|c|}{\textbf{Dispatch Cost (k\$)}} & 1,132.97 & 1,131.41 & 1,280.27& 1,278.49 \\
\hline
\multicolumn{2}{|c|}{\textbf{Total Cost (k\$)}} & 1,877.59 & 1,876.81 & 2,065.35& 2,060.04\\
\hline
\multirow{3}{1.8cm}{\centering \textbf{Maintenance Schedule}}&\textbf{G8} & 5-16 & 5-16 & 5-16 & 2-13 \\
\cline{2-6}
~&\textbf{G10} & 50-61 & 50-61 & 50-61 & 50-61 \\
\cline{2-6}
~&\textbf{G24} & 130-153 & 145-168 & 130-153 & 145-168 \\
\hline
\end{tabular}
\label{tab2}
\end{table}

As shown in Table \ref{tab2}, when uncertainty is ignored, the J-UMUC strategy reduces the total cost from 1,877.597 k\$ in Case 1 to 1,876.813 k\$ in Case 2. This reduction is achieved by adjusting the UM schedule and optimizing the UC method. Note that both cases 3 and 4 yield higher UC and dispatch costs than those of cases 1 and 2. This is because, by reserving more units while considering uncertainty, some economic efficiency is sacrificed to ensure robustness, which guarantees the safe operation of the system. Compared to Case 3, Case 4 proves to be more economical through the J-UMUC strategy. In this scenario, the maintenance schedule for units G8 and G24 is adjusted to periods 2-13 and 145-168, respectively, allowing for greater flexibility in UC during the remaining periods. This adaptability enables the system to respond effectively to severe scenarios within the uncertainty set.

\begin{figure}[b]
\vspace{-15pt}
\centerline{\includegraphics[width=3.6in]{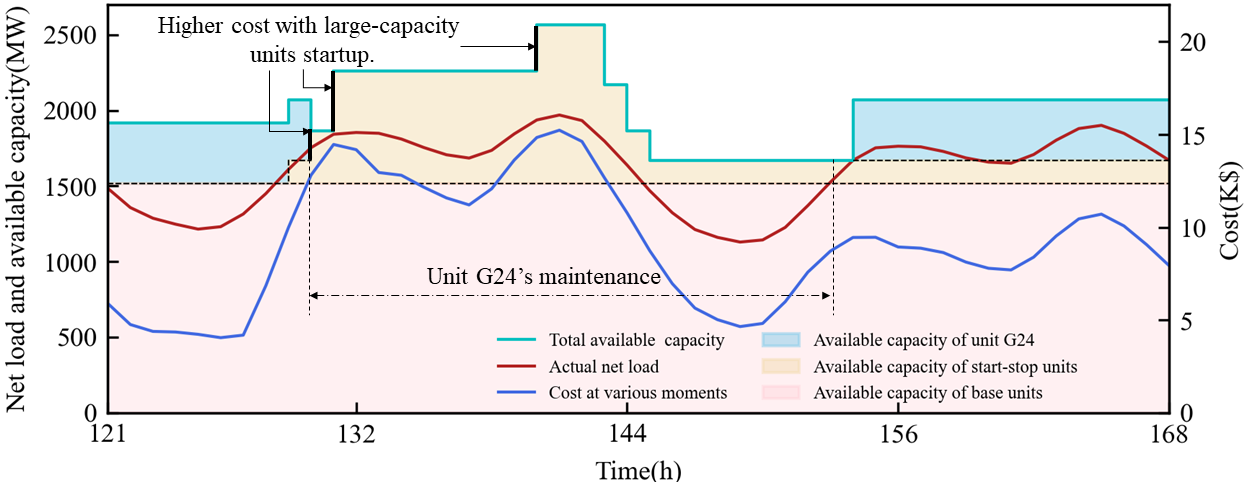}}
\centerline{(a)}
\centerline{\includegraphics[width=3.6in]{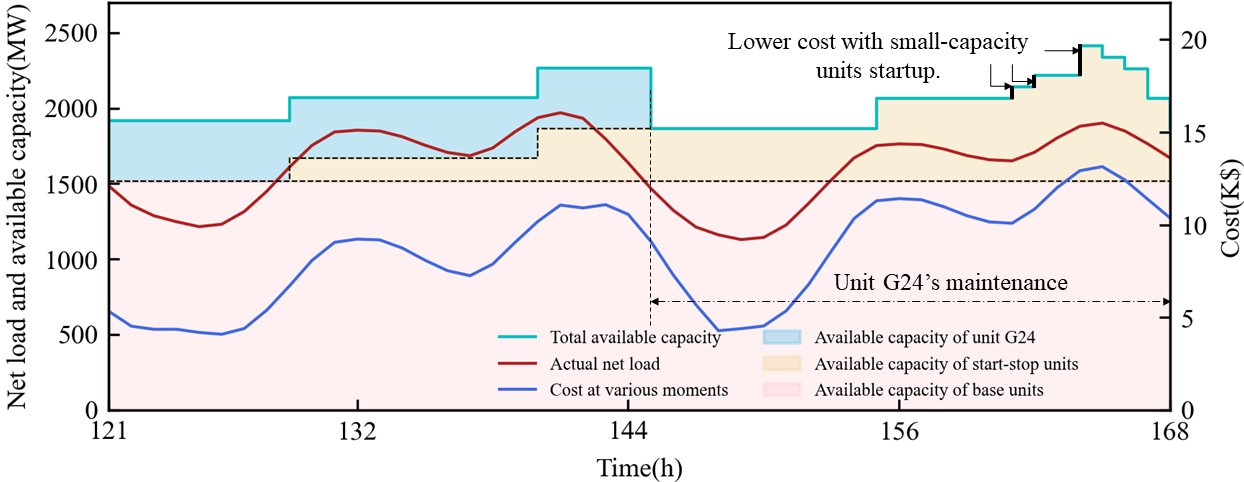}}
\centerline{(b)}
\caption{The net load curve and system available generating capacity around G24's maintenance. (a) Case 3. (b) Case 4.}
\label{fig2}
\end{figure}

To better illustrate the advantages of the J-UMUC strategy, we analyze the UM of unit G24 in cases 3 and 4. As shown in Fig. \ref{fig2}, the total available capacity includes the available capacity of unit G24, start-stop units, and base units remaining operational all the time. It can be seen that in case 3, unit G24 cannot startup during the original maintenance period, thus requiring a large capacity of start-stop units to meet the high load demand, resulting in higher generation costs. However, in case 4, postponing maintenance to a period of lower net load keeps unit G24 on to meet high load demand, and requires a smaller capacity of start-stop units during the adjusted maintenance period, making it a more economical choice.

\vspace{-5pt}
\subsection{Analysis of the Solution Algorithm}

Table \ref{tab3} compares the efficiency and accuracy of the $\text{C\&CG}$ method and the $\text{i-C\&CG}$ method under different parameters, with $\delta$ set at 0.2\%, $\tilde{\delta}$ set at 0.15\%, and the optimal gap in $\text{C\&CG}$ set at 0.01\%. Each case is calculated ten times, and the average value is taken. The results show that the $\text{i-C\&CG}$ method achieves an overall improvement in solving efficiency while maintaining an acceptable error margin. 

When $\epsilon_{MP}$ is large (e.g., $> {\delta}$), a larger $\alpha$ results in a slower solution rate. This occurs because a larger $\alpha$ causes $\epsilon_{MP}$ to shrink more slowly and tends to enlarge the scenario set in each iteration, thereby limiting solution efficiency. Conversely, when $\epsilon_{MP}$ is comparable to $\delta$, the solution rate improves, as the method is more inclined to iterate $\epsilon_{MP}$ rather than adding new scenarios. In this situation, a larger $\alpha$ can facilitate the convergence, resulting in slightly shorter solution times. For instance, when $\epsilon_{MP}=0.1\%$ and $\alpha=0.9$, the fastest solution time decreased from 2,061.5 seconds with $\text{C\&CG}$ to 930.2 seconds with $\text{i-C\&CG}$, achieving an improvement of 54.88\%. 

\begin{table}[t]
\caption{Algorithm Performance under different parameters}
\vspace{-20pt}
\begin{center}
\begin{tabular}{|c|m{0.8cm}|m{0.8cm}|c|c|m{0.8cm}|}
\hline
\multirow{3}{1cm}{\centering \textbf{Method}}&\multicolumn{2}{|c|}{\textbf{i-C\&CG }}
&\multicolumn{2}{|c|}{\textbf{i-C\&CG}}&\multirow{3}*{\textbf{C\&CG}}\\

~&\multicolumn{2}{|c|}{\textbf{($\epsilon_{MP}$=0.8\%)}}
&\multicolumn{2}{|c|}{\textbf{($\epsilon_{MP}$=0.1\%)}}&~\\
\cline{2-5}
~&\textbf{$\alpha$=0.9} & \textbf{$\alpha$=0.2} & \textbf{$\alpha$=0.9} & \textbf{$\alpha$=0.2} &~ \\
\hline
\textbf{Total Cost (k\$)} & 2,058.20 & 2,058.90 & 2,060.04& 2,060.04 & 2,060.12\\
\hline
\textbf{Solution Time (s)} & 1,986.78 & 1,680.62 & 930.24& 945.55 & 2,061.53\\
\hline
\end{tabular}
\label{tab3}
\end{center}
\vspace{-20pt}
\end{table}
\vspace{-2pt}
\section{Conclusion}

In this paper, we propose a two-stage adaptive robust optimization model for the J-UMUC strategy under the uncertainties of wind power and load demand. The first stage focuses on the J-UMUC schedule, while the second stage addresses economic dispatch. To solve this model, we propose an $\text{i-C\&CG}$ algorithm nested OA approach. 
The results from various cases demonstrates that the proposed method yields a more effective scheduling strategy under source-load uncertainty, showcasing significant cost-effectiveness and adaptability. Furthermore, we validate the efficiency advantage of the $\text{i-C\&CG}$ algorithm through algorithm parameter settings and comparisons with the standard $\text{C\&CG}$ method.

\vspace{-5pt}
\bibliography{ref}

\begin{thebibliography}{10}

\bibitem{lu1}
B.~Lindner, R.~Brits, J.~{van Vuuren}, and J.~Bekker, ``Tradeoffs between levelling the reserve margin and minimising production cost in generator maintenance scheduling for regulated power systems,'' {\em Int. J. Elec. Power}, vol.~101, pp.~458--471, 2018.

\bibitem{he1}
L.~Yang, G.~Li, Z.~Zhang, and X.~Ma, ``Operations \& maintenance optimization of wind turbines integrating wind and aging information,'' {\em IEEE Trans. Sustain. Energy}, vol.~12, no.~1, pp.~211--221, 2021.

\bibitem{lu3}
E.~Du, N.~Zhang, C.~Kang, and Q.~Xia, ``A high-efficiency network-constrained clustered unit commitment model for power system planning studies,'' {\em IEEE Trans. Power Syst.}, vol.~34, no.~4, pp.~2498--2508, 2019.

\bibitem{lu2}
M.~Yildirim, N.~Z. Gebraeel, and X.~A. Sun, ``Leveraging predictive analytics to control and coordinate operations, asset loading, and maintenance,'' {\em IEEE Trans. Power Syst.}, vol.~34, no.~6, pp.~4279--4290, 2019.

\bibitem{lu4}
P.~Ramanan, M.~Yildirim, N.~Gebraeel, and E.~Chow, ``Large-scale maintenance and unit commitment: A decentralized subgradient approach,'' {\em IEEE Trans. Power Syst.}, vol.~37, no.~1, pp.~237--248, 2022.

\bibitem{lu9}
Y.~Wang, Z.~Li, M.~Shahidehpour, L.~Wu, C.~X. Guo, and B.~Zhu, ``Stochastic co-optimization of midterm and short-term maintenance outage scheduling considering covariates in power systems,'' {\em IEEE Trans. Power Syst.}, vol.~31, no.~6, pp.~4795--4805, 2016.

\bibitem{robust}
Y.~Danwen, Y.~Ming, Z.~Hefeng, and H.~Xueshan, ``An overview of robust optimization used for power system dispatch and decision-making,'' {\em Automation of Electric Power Systems}, vol.~40, no.~07, pp.~134--143+148, 2016.

\bibitem{lu15}
M.~Y. Tsang, K.~S. Shehadeh, and F.~E. Curtis, ``An inexact column-and-constraint generation method to solve two-stage robust optimization problems,'' {\em OPER RES LETT}, vol.~51, no.~1, pp.~92--98, 2023.

\bibitem{lu17}
D.~Bertsimas, E.~Litvinov, X.~A. Sun, J.~Zhao, and T.~Zheng, ``Adaptive robust optimization for the security constrained unit commitment problem,'' {\em IEEE Trans. Power Syst.}, vol.~28, no.~1, pp.~52--63, 2013.

\bibitem{lu16}
C.~Grigg {\em et~al.}, ``The ieee reliability test system-1996. a report prepared by the reliability test system task force of the application of probability methods subcommittee,'' {\em IEEE Trans. Power Syst.}, vol.~14, no.~3, pp.~1010--1020, 1999.

\end{thebibliography}

\end{document}